\documentclass[twocolumn,aps,superscriptaddress,showpacs]{revtex4}
\usepackage{amssymb}
\usepackage{amsmath,bm}
\usepackage{graphicx}
\usepackage[normalem]{ulem}
\usepackage[dvips]{color}

\renewcommand\sout{\bgroup \color{red} \ULdepth=-.5ex \ULset}

\begin{document}

\title{Isospin splitting of nucleon effective mass from giant resonances in $^{208}$Pb}
\author{Zhen Zhang}
\affiliation{Department of Physics and Astronomy and Shanghai Key Laboratory for
Particle Physics and Cosmology, Shanghai Jiao Tong University, Shanghai 200240, China}
\author{Lie-Wen Chen\footnote{%
Corresponding author (email: lwchen$@$sjtu.edu.cn)}}
\affiliation{Department of Physics and Astronomy and Shanghai Key Laboratory for
Particle Physics and Cosmology, Shanghai Jiao Tong University, Shanghai 200240, China}
\affiliation{Center of Theoretical Nuclear Physics, National Laboratory of Heavy Ion
Accelerator, Lanzhou 730000, China}
\date{\today}

\begin{abstract}
Based on mean field calculations with Skyrme interactions,
we extract a constraint on the isovector effective mass in nuclear matter at
saturation density $\rho_0$, i.e., $m_{v}^{\ast}(\rho_0)=(0.77\pm0.03) m$ by
combining the experimental data of the centroid energy of the isovector giant
dipole resonance (IVGDR) and the electric dipole polarizability $\alpha_{\mathrm{D}}$
in $^{208}$Pb. Meanwhile, the isoscalar effective mass at $\rho_0$ is determined to be
$m_{s}^{\ast}(\rho_0)=(0.91\pm0.05) m$ by analyzing the measured excitation
energy of the isoscalar giant quadrupole resonance (ISGQR) in $^{208}$Pb.
From the constrained $m_{s}^{\ast}(\rho_0)$ and $m_{v}^{\ast}(\rho_0)$,
we obtain the isospin splitting of nucleon effective mass in asymmetric nuclear
matter of isospin asymmetry $\delta$ at $\rho_0$ as
$[m_n^{\ast}(\rho_0,\delta)-m_p^{\ast}(\rho_0,\delta)]/m = \Delta m^*_1(\rho_0) \delta + O(\delta^3)$
with the linear isospin splitting coefficient $\Delta m^*_1(\rho_0) = 0.33\pm0.16$.
We notice that using the recently corrected data on the $\alpha_{\mathrm{D}}$
in $^{208}$Pb with the contribution of the quasideuteron effect
subtracted slightly enhances the isovector effective mass to
$m_{v}^{\ast}(\rho_0)=(0.80\pm0.03) m$ and reduces the linear isospin splitting
coefficient to $\Delta m^*_1(\rho_0) = 0.27\pm0.15$.
Furthermore, the constraints on $m_{v}^{\ast}(\rho)$, $m_{s}^{\ast}(\rho)$ and
$\Delta m^*_1(\rho)$ at other densities are obtained from the
similar analyses and we find that the $\Delta m^*_1(\rho)$ increases with the density.

\end{abstract}

\pacs{21.65.-f, 24.30.Cz, 21.60.Jz, 21.30.Fe}
\maketitle

\section{Introduction}
Nucleon effective mass, which is usually introduced to characterize the
dynamical properties for the propagation of (quasi)nucleons in nuclear
medium, is of fundamental importance in nuclear many-body physics~\cite{Jeu76,Sjo76,Neg81}.
While there exist several different kinds of nucleon effective masses in
non-relativistic and relativistic approaches~\cite{Mah85,Jam89,Chen07,LCK08},
we shall focus in this work on the total nucleon effective mass used typically
in the non-relativistic approach, which measures the
momentum dependence (or equivalently energy dependence by assuming an on-shell
dispersion relation) of the nucleon single-particle potential in nuclear medium. In isospin
asymmetric nuclear matter, neutrons and protons may feel different single-particle
potentials which can then lead to the isospin splitting of nucleon effective mass,
i.e., $m^{\ast}_{n-p}\equiv (m_n^{\ast}-m_p^{\ast})/m$.
The isospin splitting of nucleon effective mass may have a profound impact
on various physical phenomena and quantities in nuclear physics, astrophysics
and cosmology~\cite{Mei07,LiChen15}, such as the properties of mirror nuclei~\cite{Nol69},
transport properties of asymmetric nuclear matter~\cite{Li04,Li05,Riz05,Gio10,Fen12,Zha14,Xie14,Beh11,XuJ15a,XuJ15b},
neutrino emission in neutron stars~\cite{Bal14}, and the primordial nucleosynthesis
in the early universe~\cite{Ste06}.
The isospin splitting of nucleon effective mass is also related to the
momentum dependence of the nuclear isovector (symmetry) potential in nuclear medium~\cite{LiBA04}
and thus the nuclear symmetry energy~\cite{XuC10,XuC11,ChenR12,CaiBJ12,ChenLW14} which is
of critical importance for many issues of both nuclear physics and astrophysics but
remains largely uncertain. A most recent review on the isospin splitting of nucleon
effective mass as well as its relation to the symmetry energy and symmetry potential
can be found in Re.~\cite{LiChen15}.

Theoretical studies based on either microscopic many-body theories or
phenomenological approaches have thus far given widely divergent predictions on
$m^{\ast}_{n-p}$. For example, non-relativistic Brueckner-Hartree-Fock and
relativistic Dirac-Brueckner-Hartree-Fock calculations indicate
$m^{\ast}_{n-p}>0$~\cite{Zuo99,Ma04,Dal05} in neutron-rich matter, while
relativistic mean field, Skyrme-Hartree-Fock (SHF) and Gogny-Hartree-Fock
models predict either $m^{\ast}_{n-p}>0$ or
$m^{\ast}_{n-p}<0$~\cite{Chen07,Bar05,Lon06,Gor09,Ou11,Dut12,ChenR12,Sel14},
depending on the interactions.
During the last several years, significant progress has been made in determining
the isospin splitting of nucleon effective mass by analyzing experimental data~\cite{LiChen15}.
However, there is still no quantitatively and even
qualitatively consensus on the behavior of $m^{\ast}_{n-p}$ in asymmetric nuclear
matter. For example, while the optical model analyses of nucleon-nucleus scattering
data~\cite{XuC10,LiXH15} favor $m^{\ast}_{n-p}>0$ in neutron-rich matter at $\rho_0$,
the transport model analysis on the double $n/p$ ratio in heavy ion collisions
seems to suggest the opposite conclusion~\cite{Cou14} (but see Ref.~\cite{Kon15}).
Therefore, any new and independent constraints on the isospin splitting of
nucleon effective mass are extremely helpful for understanding the issue on the
behavior of $m^{\ast}_{n-p}$ in asymmetric nuclear matter.

Nuclear giant resonances provide an important approach to determine nucleon effective
mass. It has been well established that the excitation energy $E_x$ of the isoscalar
giant quadrupole resonance (ISGQR) in finite nuclei is related to the
isoscalar effective mass $m_{s}^{\ast}(\rho)$ (nucleon effective mass in symmetric
nuclear matter) at $\rho_0$, i.e., $m_{s,0}^{\ast}$ (see, e.g., Refs.~\cite{Boh75,Boh79,Bla80,Roc13}).
A value of $m_{s,0}^{\ast}\sim 0.8m$ has been estimated by analyzing experimental
data for ISGQR excitation energy in early studies~\cite{Boh79}, and more recent
microscopic random phase approximation (RPA) calculations suggest that the ISGQR in
heavy nuclei favors $m_{s,0}^{\ast}\sim 0.9m$~\cite{Bla80,Klu09,Roc13}.
Moreover, within the RPA approach
using Skyrme interactions, the isovector effective mass $m_{v}^{\ast}(\rho)$ (i.e.,
neutron (proton) effective mass in pure proton (neutron) matter) at $\rho_0$,
i.e., $m_{v,0}^{\ast}$, is closely related to the enhancement factor $\kappa$ of
the energy weighted sum rule (EWSR) $m_{1}$ in the isovector giant dipole resonance
(IVGDR)~\cite{Cha97,Les06,Klu09}. Unfortunately, while the peak of the IVGDR
strength function has been well located for a number of nuclei by photoabsorbtion
measurements~\cite{Die88}, neither the $m_{1}$ nor the $\kappa$ has been
accurately determined.
Furthermore, the detailed relation between $m_{v,0}^{\ast}$ and
$m_{1}$ or $\kappa$ has not yet systematically investigated for different nuclei.
Therefore, the $m_{v,0}^{\ast}$ has so far not yet been properly constrained.

The properties of the heavy doubly magic nucleus $^{208}$Pb, especially
including various kinds of giant resonances, have been well researched. In the
present work, we mainly study how the ISGQR and IVGDR of $^{208}$Pb constrain
the $m^{\ast}_{n-p}$ in neutron-rich matter.
Thanks to the recent high resolution measurement for the electric dipole
polarizability $\alpha_{\mathrm{D}}$ in $^{208}$Pb, which is determined
by the inverse energy weighted sum rule $m_{-1}$ of the IVGDR, at the Research
Center for Nuclear Physics (RCNP)~\cite{Tam11}, in this work, we deduce the $m_{1}$
of the IVGDR in $^{208}$Pb from the experimental value of the IVGDR centroid
energy $E_{-1}=\sqrt{m_1/m_{-1}}$~\cite{Die88}. Using the RPA calculations
with a number of representative Skyrme interactions, we establish the
detailed relations between $m_{s,0}^{\ast}$ ($m_{v,0}^{\ast}$) and the $E_x$
of ISGQR ($m_{1}$) in $^{208}$Pb, and
then extract relatively accurate constraints on $m_{s}^{\ast}$ and
$m_{v}^{\ast}$ from the ISGQR excitation energy $E_x$ and the $m_{1}$ of the
IVGDR in $^{208}$Pb, respectively.
Within the SHF model, we show that the $m^{\ast}_{n-p}$ is
completely determined by the $m_{s}^{\ast}$ and $m_{v}^{\ast}$, and thus
we can obtain constraints on the
$m^{\ast}_{n-p}$, which is the main motivation of the present work.
For the first time, our results indicate that the data on the giant resonances in
$^{208}$Pb definitely favor $m^{\ast}_{n-p}>0$ in neutron-rich matter, which would be
very helpful to pin down the isospin splitting of nucleon effective mass.

\section{Model and method}

\subsection{Nucleon effective mass in Skyrme-Hartree-Fock approach}

In non-relativistic approaches, the effective mass $m_q^*$ of a
nucleon $q$ (n or p) in asymmetric nuclear matter with density $\rho $ and
isospin asymmetry $\delta=(\rho _{n}-\rho _{p})/(\rho _{p}+\rho _{n})$ can be calculated as~\cite{Jeu76}
\begin{eqnarray}\label{effmass}
\frac{m_q^{\ast }(\rho,\delta)}{m_q}=\left[ 1+\frac{m_q}{k}\frac{dU_q(k,\epsilon _q(k,\rho,\delta),\rho,\delta)}{dk}\Bigg|_{k_F^q}\right]
^{-1},\label{mstar}
\end{eqnarray}
where $m_q$ represents the mass of neutrons or protons in free-space ($m_q = m$ is assumed in this work),
$k_F^q$ is the neutron/proton Fermi momentum, $U_q$ is the single-nucleon potential, and $\epsilon _q$ is
the nucleon single-particle energy satisfying the following dispersion relation
\begin{eqnarray}
\epsilon _q(k,\rho,\delta)=\frac{k^{2}}{2m_q}+U_q(k,\epsilon_q(k),\rho,\delta).
\end{eqnarray}

In this work, we  use the standard Skyrme interaction with a zero-range and
velocity-dependent form as~\cite{Cha97}
\begin{eqnarray}
V_{12}(\mathbf{R},\mathbf{r}) &=&t_{0}(1+x_{0}P_{\sigma })\delta (\mathbf{r})
\notag \\
&+&\frac{1}{6}t_{3}(1+x_{3}P_{\sigma })\rho ^{\sigma }(\mathbf{R})\delta (%
\mathbf{r})  \notag \\
&+&\frac{1}{2}t_{1}(1+x_{1}P_{\sigma })(\mathbf{K}^{^{\prime }2}\delta (\mathbf{r}%
)+\delta (\mathbf{r})\mathbf{K}^{2})  \notag \\
&+&t_{2}(1+x_{2}P_{\sigma })\mathbf{K}^{^{\prime }}\cdot \delta (\mathbf{r})%
\mathbf{K}  \notag \\
&\mathbf{+}&iW_{0}(\mathbf{\sigma }_{1}+\mathbf{\sigma }_{2})\cdot \lbrack
\mathbf{K}^{^{\prime }}\times \delta (\mathbf{r})\mathbf{K]},  \label{V12Sky}
\end{eqnarray}%
with $\mathbf{r}=\mathbf{r}_{1}-\mathbf{r}_{2}$ and $\mathbf{R}=(\mathbf{r}%
_{1}+\mathbf{r}_{2})/2$. In the above expression, the relative momentum
operators $\mathbf{K}=(\mathbf{\nabla }_{1}-\mathbf{\nabla }_{2})/2i$ and $%
\mathbf{K}^{\prime }=-(\mathbf{\nabla }_{1}-\mathbf{\nabla }_{2})/2i$ act on
the wave function on the right and left, respectively. The quantities $%
P_{\sigma }$ and $\sigma _{i}$ denote, respectively, the spin exchange
operator and Pauli spin matrices. In the following, several Skyrme interactions
with nonstandard spin-orbit term~\cite{Rei95} are also employed, but the
spin-orbit term is irrelevant to the expressions of nucleon effective mass.

Within the standard SHF approach, the nucleon effective
mass in asymmetric nuclear matter with density $\rho $ and
isospin asymmetry $\delta$ can be expressed as~\cite{Cha97}
\begin{eqnarray}
\label{Eq:EM}
\frac{\hbar ^2}{2m_{q}^{\ast}(\rho,\delta)}&=&\frac{\hbar ^2}{2m}
+\frac{1}{4}t_1\left[\left(1+\frac{1}{2}x_1\right)\rho
-\left(\frac{1}{2}+x_1\right)\rho_q\right] \notag \\
&+&\frac{1}{4}t_2\left[\left(1+\frac{1}{2}x_2\right)\rho
+\left(\frac{1}{2}+x_2\right)\rho_q\right].
\end{eqnarray}
By setting $\rho_q=\rho/2$ in Eq.~(\ref{Eq:EM}), the isoscalar effective
mass can then be obtained as~\cite{Cha97}
\begin{equation}
\label{Eq:ISEM}
  \frac{\hbar^2}{2m_{s}^{\ast}(\rho)}=\frac{\hbar^2}{2m}+\frac{3}{16}t_1\rho+\frac{1}{16}t_2(4x_2+5)\rho.
\end{equation}
The isovector effective mass, which corresponds to the proton (neutron) effective
mass in pure neutron (proton) matter, can be obtained with $\rho_q=0$ in Eq.~(\ref{Eq:EM}) as~\cite{Cha97}
\begin{equation}
\label{Eq:IVEM}
   \frac{\hbar^2}{2m_{v}^*(\rho)}=\frac{\hbar^2}{2m}+\frac{1}{8}t_1(x_1+2)\rho+\frac{1}{8}t_2(x_2+2)\rho.
\end{equation}

From Eqs.~(\ref{Eq:EM}), (\ref{Eq:ISEM}) and (\ref{Eq:IVEM}), one can
obtain the isospin splitting of nucleon effective mass, i.e.,
\begin{eqnarray}
\label{Eq:DEMNP}
m^{\ast}_{n-p}(\rho,\delta)&\equiv &\frac{m_{n}^{\ast}-m_{p}^{\ast}}{m}=2\frac{m_{s}^{\ast}}{m}\sum_{n=1}^{\infty} \left(\frac{m_{s}^{\ast}-m_{v}^{\ast}}{m_{v}^{\ast}}\delta\right)^{2n-1} \notag \\
&=& \sum_{n=1}^{\infty} \Delta m^*_{2n-1}(\rho)\delta^{2n-1},
\end{eqnarray}
where the isospin splitting coefficients $\Delta m^*_{2n-1}(\rho)$ can be
expressed as
\begin{eqnarray}
\label{DmIso}
\Delta m^*_{2n-1}(\rho) = 2\frac{m_{s}^{\ast}}{m} \left(\frac{m_{s}^{\ast}}{m_{v}^{\ast}}-1\right)^{2n-1}.
\end{eqnarray}
The above expressions reveal that, within the SHF model, the $m^{\ast}_{n-p}$
is completely determined by the $m_{s}^{\ast}$ and $m_v^{\ast}$, and the sign
of $m^{\ast}_{n-p}$ in neutron-rich matter is the same as that of
$m_{s}^{\ast} - m_v^{\ast}$.

\subsection{Random-phase approximation and nuclear giant resonances}

The random-phase approximation~\cite{Rin80} provides a successful microscopic
approach to study giant resonance observables in finite nuclei.
Within the framework of RPA theory, for a given excitation operator $\hat{F}_{JM}$,
the reduced transition probability from RPA ground state $|\tilde{0}\rangle $
to RPA excitation state $|\nu \rangle  $ is given by:
\begin{equation}
\begin{split}
B(EJ:\tilde{0}\rightarrow|\nu\rangle)&=|\langle\nu||\hat{F}_{J}||\tilde{0}\rangle|^2\\
&=\left|\sum_{mi}\left( X_{mi}^{\nu}+Y_{mi}^{\nu}\right) |\langle m||\hat{F}_{J}||i\rangle\right| ^2
\end{split},
\end{equation}
where $m (i)$ denotes the unoccupied (occupied) single nucleon state;
$\langle m||\hat{F}_{J}||i\rangle$ is the reduced matrix element of $\hat{F}_{JM}$;
and $X_{mi}^{\nu}$ and $Y_{mi}^{\nu}$ are the RPA amplitudes. The strength
function then can be calculated as:
\begin{equation}
S(E)=\sum_{\nu}|\langle\nu\Vert\hat{F}_J\Vert\tilde{0}\rangle|^2\delta(E-E_{\nu}),
\end{equation}
where $E_{\nu}$ is the energy of RPA excitation state $|\nu\rangle$. Thus the
moments of strength function can be obtained as:
\begin{equation}
m_k=\int dE E^kS(E)=\sum_{\nu}|\langle\nu\Vert\hat{F}_J\Vert\tilde{0}\rangle|^2E_{\nu}^k.
\end{equation}
For the IVGDR and IVGQR that we are interested in here, the excitation
operators are defined as:
\begin{eqnarray}
\hat{F}_{1M} &=& \frac{N}{A}\sum^Z_{i=1}r_iY_{\text{1M}}(\hat{r}_i)-\frac{Z}{A}\sum^N_{i=1}r_iY_{\text{1M}}(\hat{r}_i), \\
\hat{F}_{2M} &=& \sum_{i=1}^{A}r_i^2Y_{2M}(\hat{r}_i),
\end{eqnarray}
where $Z$, $N$ and $A$ are proton, neutron and mass number, respectively;
$r_i$ is the nucleon's radial coordinate; $Y_{\text{1M}}(\hat{r_i})$ and  $Y_{\text{2M}}(\hat{r_i})$ are the corresponding
spherical harmonic function.

\subsection{Nucleon effective mass and nuclear giant resonances}
It is well known that the isoscalar effective mass at saturation density,
i.e., $m_{s,0}^{\ast}$, is intimately related to the excitation energy of
the ISGQR in finite nuclei. In the harmonic oscillator model, the ISGQR
energy is~\cite{Boh75,Roc13}
\begin{equation}
\label{Eq:EGQRQHO}
E_{x}=\sqrt{\frac{2m}{m_{s,0}^{\ast}}}\hbar\omega_0,
\end{equation}
where $\hbar\omega_0$ is the frequency of the harmonic oscillator. This semiempirical
expression reveals the correlation between the ISGQR excitation energy and the isoscalar
effective mass $m_{s,0}^{\ast}$, which has been also confirmed by microscopic
calculations~\cite{Bla80,Roc13}.

Meanwhile, the isovector effective mass at saturation density $m_{v,0}^{\ast}$ is
correlated with the energy weighted sum rule $m_{1}$ of the IVGDR~\cite{Col13}, i.e.
\begin{equation}
\label{Eq:Sm1}
m_{1}=\frac{9}{4\pi}\frac{\hbar ^2}{2m}\frac{NZ}{A}(1+\kappa),
\end{equation}
where $\kappa$ is the enhancement factor reflecting the deviation from the
Thomas-Reiche-Kuhn sum rule~\cite{Har01} (e.g., due to the exchange and momentum dependent force).
Within the Skyrme-RPA approach, $\kappa$ is given by~\cite{Cha97,Col13}
\begin{eqnarray}
\label{Eq:Kappa}
\kappa &=& \frac{2m}{\hbar ^2}\frac{A}{4NZ}\int \rho_n (r) \rho_p(r) d^3r
\notag \\
&&\cdot \left[ t_1\left( 1+\frac{x_1}{2}\right) +t_2\left(1+\frac{x_2}{2}\right)
\right] .
\end{eqnarray}
Substituting Eqs.~(\ref{Eq:IVEM}) and (\ref{Eq:Kappa}) into Eq.~(\ref{Eq:Sm1}) leads to
\begin{eqnarray}
m_{1}&=&\frac{9}{4\pi}\frac{\hbar ^2}{2m}\frac{NZ}{A} \notag\\
& & \cdot\left[ 1+\frac{A}{NZ}\left(\frac{m}{m_{v,0}^{\ast}}-1\right)
\frac{\int \rho_n(r)\rho_p(r)d^3r }{\rho_0}\right],
\end{eqnarray}
which suggests that the EWSR $m_{1}$ (and thus $\kappa$) of the IVGDR is proportional to
$\left(m_{v,0}^{\ast}/m \right)^{-1}$ for a fixed nucleus. In particular, by assuming
$\rho_n=\rho_p=\rho_0/2$, one then obtains the following approximate expressions~\cite{Cha97}
\begin{equation}
\label{Eq:m1mv}
m_{1} \approx \frac{9}{4\pi}\frac{\hbar ^2}{2m}\frac{NZ}{A}\left(
\frac{m_{v,0}^{\ast}}{m}\right)^{-1},
\end{equation}
and
\begin{equation}
\label{Eq:mvk}
m_{v,0}^{\ast}/m \approx 1/(1+\kappa).
\end{equation}

\section{Results and discussions}
To study the correlation between the nucleon effective mass and the giant resonance
observables, we select $50$ representative Skyrme interactions~\cite{Zha13,Dut12,Roc12}
(i.e., BSk1, BSk2, BSk5, BSk6, BSk13, Es, Gs,
KDE, KDE0v1, MSk7, MSL0, MSL1, NRAPR, Rs, SAMi, SGI, SGII, SK255, SK272,
SKa, SkI3, SkM, SkMP, SkM$^{\ast}$, SkP, SkS1, SkS2, SkS3, SkS4, SkSC15, SkT7,
SkT8, SkT9, SKX, SKXce, SKXm, Skxs15, Skxs20, SLy4, SLy5, SLy10, SV-K241, v070,
v075, v080, v090, v105, v110, Zs, Zs$^{\ast}$).
The corresponding ISGQR excitation energies and EWSRs of the IVGDR in $^{208}$Pb are
calculated by using the Skyrme-RPA program by Col$\grave{\text{o}}$ {\it et al}~\cite{Col13}.

\begin{figure}[tbp]
\includegraphics[scale=0.35]{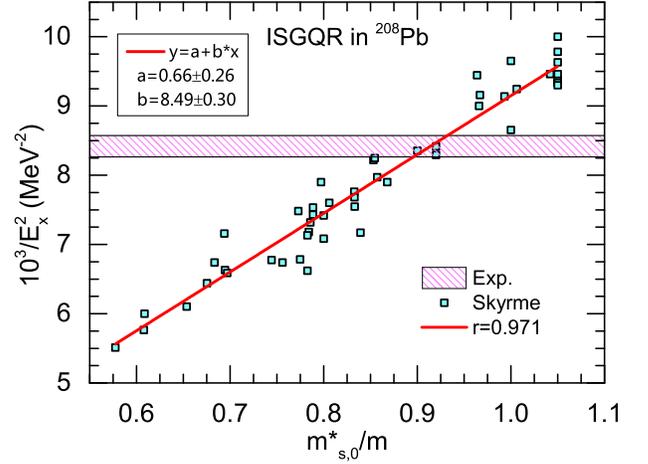}
\caption{(Color online) $10^3/E_x^2$ in $^{208}$Pb vs $m_{s,0}^{\ast}/m$ predicted
by a large number ($50$) of Skyrme interactions. The linear fit gives
${10^3}/{E_x^2}=(0.66\pm0.26)+(8.49\pm0.30)\left({m_{s,0}^{\ast}}/{m}\right)$
with the Pearson correlation coefficient being 0.971.
The hatched band corresponds the weighted averages of the experimental
values for the ISGQR excitation energy in $^{208}$Pb, $E_x=10.9\pm0.1$ MeV~\cite{Roc13}.}
\label{Exms}
\end{figure}

In the calculation of the ISGQR excitation energy $E_x$, we smear out
the strength function with Lorentzian functions with a width $1$ MeV. We note
that varying the width has little influence on the peak energy. The
obtained data-to-data relations between $ 10^3/E_x^2$ in $^{208}$Pb and $m_{s,0}/m$
predicted by the chosen $50$ Skyrme interactions are displayed in Fig.~\ref{Exms}.
Also included in Fig.~\ref{Exms} is the linear fit together with the
corresponding Pearson correlation coefficient $r$. As expected from the semiempirical
relation Eq.~(\ref{Eq:EGQRQHO}), one can see that a strong linear correlation
exists between $1/E_x^2$ and $m_{s,0}^{\ast}/m$ with the coefficient $r$ as
large as $0.971$. And the linear fit gives
\begin{equation}
\label{Eq:FitExMs}
\frac{10^3}{E_x^2}=(0.66\pm0.26)+(8.49\pm0.30)\left(\frac{m_{s,0}^{\ast}}{m}\right),
\end{equation}
where the $E_x$ is in MeV.

In the present work, we invoke the weighted average of experimental values
for the ISGQR energy in $^{208}$Pb, i.e., $E_x=10.9\pm0.1$ MeV~\cite{Roc13},
which is shown as the hatched band in Fig.~\ref{Exms}. Combining this weighted
average and Eq.~(\ref{Eq:FitExMs}), we extract the isoscalar effective mass
at saturation density as
\begin{equation}
\frac{m_{s,0}^{\ast}}{m}=0.91\pm0.05 .
\end{equation}
Here the error is obtained from the propagation of the experimental uncertainty
of $E_x$ and parameter errors in the linear fit.
This constraint is consistent with the result $m_{s,0}^{\ast}\approx 0.8-0.9m$
obtained from analyzing the ISGQR of Nd and Sm isotopes~\cite{Yos13}, and naturally
confirms the empirical value of $m_{s,0}^{\ast}\sim0.9m$ predicted by some Skyrme
interactions which are obtained by fitting the experimental data of the ISGQR
excitation energy in finite nuclei~\cite{Bla80,Klu09}. It is also in good
agreement with the result of $m_{s,0}^{\ast}\sim0.92m$ from the extended
Brueckner-Hartree-Fock calculation with realistic nucleonic forces~\cite{Zuo99}.

\begin{figure}[btp]
\includegraphics[scale=0.35]{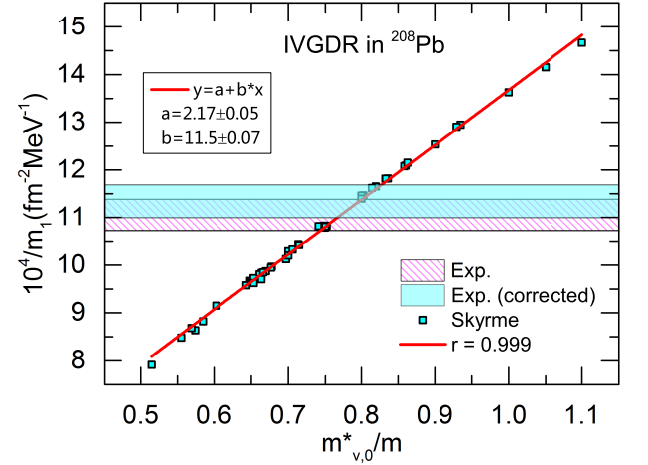}
\caption{(Color online) $10^4/m_1$ in $^{208}$Pb vs $m_{v,0}^{\ast}/m$
predicted by a large number ($50$) of Skyrme interactions. The linear
fit gives ${10^4}/{m_1} = (2.17\pm0.05)+(11.5\pm0.07){m_{v,0}^{\ast}}/{m}$
with the Pearson correlation coefficient being 0.999.
The hatched band (cyan band) corresponds to the (corrected)
experimental value of the EWSR $m_1$ of the IVGDR in $^{208}$Pb (see text for the details).}
\label{m1mv}
\end{figure}

For the IVGDR, we use the chosen $50$ Skyrme interactions to evaluate the
$m_{1}$ of the IVGDR in $^{208}$Pb with energy up to $130$ MeV. Similarly, in
Fig. \ref{m1mv}, we plot the data-to-data relations between $10^4/m_1$ and
$m_{v,0}^{\ast}/m$ as well as the linear fit and Pearson correlation coefficient
$r$. It is clearly shown that an excellent linear correlation exists between
$1/m_1$ and $m_{v,0}^{\ast}/m$, and the linear fit gives
\begin{equation}
\label{Eq:FitM1Mv}
\frac{10^4}{m_1} = (2.17\pm0.05)+(11.5\pm0.07)\frac{m_{v,0}^{\ast}}{m},
\end{equation}
where the $m_1$ is in MeV$\cdot\mathrm{fm}^2$. Experimentally, the centroid
energy of the IVGDR, i.e., $E_{-1}=\sqrt{m_1/m_{-1}}$, in $^{208}$Pb has been
well determined from photoabsorbtion measurements, i.e., $E_{-1}=13.46$
MeV~\cite{Die88}, and the inverse energy weighted sum rule $m_{-1}$
can be obtained from the experimental value of the electric dipole polarizability
measured at RCNP, i.e., $\alpha_{\mathrm{D}}=20.1\pm0.6~\mathrm{fm}^{3}$\cite{Tam11},
through the following simple relation
\begin{equation}
m_{-1}=\frac{9}{8\pi e^2}\alpha_{\mathrm{D}}.
\end{equation}
The experimental values of $E_{-1}$ and $m_{-1}$ together thus give
$m_{1}=905.60\pm27.03~\mathrm{MeV}\cdot\mathrm{fm}^2$. Therefore, one can
constrain isovector effective mass at saturation density using
Eq.~(\ref{Eq:FitM1Mv}), and the result is
\begin{equation}
\frac{m_{v,0}^{\ast}}{m}=0.77 \pm 0.03.
\end{equation}
One can see that our constraint is rather accurate and well consistent with
the empirical value, e.g., $m_{v,0}^{\ast}/{m}=0.90\pm0.2$ from analyses of
finite nuclei mass data~\cite{Pea01}. In addition, from the well known
relation Eq.~(\ref{Eq:Sm1}) for the EWSR, the value of the enhancement factor
$\kappa$ can be deduced as $\kappa=0.228\pm0.037$ with
$m_{1}=905.60\pm27.03~\mathrm{MeV}\cdot\mathrm{fm}^2$,
which is in very good agreement with
$\kappa=0.22\pm0.04$ reported in Ref.~\cite{Tri08} and consistent with
the estimate of $\kappa \approx 0.2 - 0.3$ in Ref.~\cite{Lip89}.
We note that using the relation $m_{v,0}^{\ast}/m \approx 1/(1+\kappa)$
(i.e., Eq.~(\ref{Eq:mvk})) leads to a little bit larger $\kappa$ as
$\kappa \approx 0.30\pm0.05$, indicating that Eq.~(\ref{Eq:mvk}) is
indeed satisfied approximately. However, one should be
cautious to use the relation $m_{v,0}^{\ast}/m \approx 1/(1+\kappa)$ for an
accurate determination on $m_{v,0}^{\ast}$ from $\kappa$, and vice verse.

From the constraints on $m_{s,0}^{\ast}$ and $m_{v,0}^{\ast}$, one can then
obtain the isospin splitting $m^*_{n-p}(\rho_0)$ according to Eq.~(\ref{DmIso}).
In particular, we obtain the first-order (linear) isospin splitting coefficient
$\Delta m^*_1(\rho)$ at $\rho_0$ as
\begin{equation}
\Delta m^*_1(\rho_0)=0.33\pm0.16,
\end{equation}
which is in very good agreement with the constraint $\Delta m^*_1(\rho_0)=0.32\pm0.15$
obtained in Ref.~\cite{XuC10} and the more recent constraint $\Delta m^*_1(\rho_0)=0.41\pm0.15$
extracted in Ref.~\cite{LiXH15} from the global optical model analysis of nucleon-nucleus
scattering data. The present result is also consistent with the
$\Delta m^*_1(\rho_0)=0.27$ obtained by analyzing various constraints on the
magnitude and density slope of the symmetry energy at $\rho_0$~\cite{LiBA13}.
The positive value of $\Delta m^*_1(\rho_0)$ further agrees with the
microscopic Brueckner calculations with realistic nuclear forces~\cite{Zuo99,Ma04,Dal05}.
In addition, it is interesting to see that the higher-order isospin splitting coefficients
are rather small and can be neglected safely. For example, the third-order isospin
splitting coefficient $\Delta m^*_3(\rho_0)$ is found to be $0.01\pm0.01$.

\begin{figure}[tbp]
\centering
\includegraphics[scale=0.5]{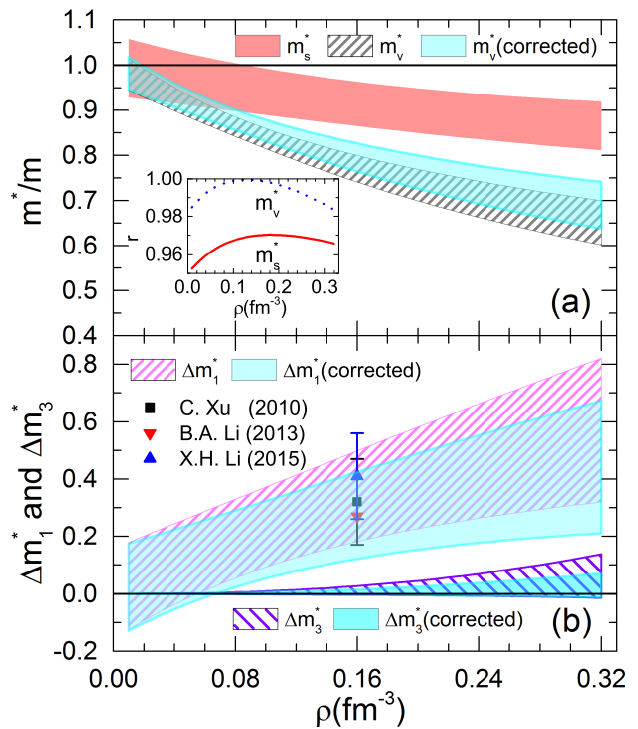}
\caption{(Color online) Panel (a): Constraints on the density dependence of the
isoscalar and isovector effective mass, $m_{s}^{\ast}$ and $m_v^{\ast}$, extracted
from the ISGQR and IVGDR in $^{208}$Pb, respectively. The inset shows the
corresponding Pearson correlation coefficient $r$ as a function of density.
Panel (b): Constraints on the density dependence of the isospin splitting coefficients
$\Delta m^*_1(\rho)$ and $\Delta m^*_3(\rho)$ obtained in this work.
The hatched bands (cyan bands) represent the results of $m_v^{\ast}$,
$\Delta m^*_1(\rho)$ and $\Delta m^*_3(\rho)$  without (with) subtracting the
contribution from the quasideuteron effect.
The $\Delta m^*_1(\rho_0)$
constraints obtained in Refs.~\cite{XuC10,LiXH15,LiBA13} are also included for comparison.
}
\label{dmrho}
\end{figure}

The above analyses are only made at saturation density $\rho_0$ and it is also interesting to
see the constraints on $m_{s}^{\ast}(\rho)$, $m_{v}^{\ast}(\rho)$ and $m^*_{n-p}(\rho)$
at other densities. Similar analyses indicate that the strong linear correlation also
exists between $1/E_x^2$ and $m_{s}^{\ast}(\rho)/m$ as well as between $1/m_1$ and
$m_{v}^{\ast}(\rho)/m$ at other densities $\rho$. Shown in Fig.~\ref{dmrho} (a) are the
constraints on the $m_{s}^{\ast}(\rho)/m$ and $m_{v}^{\ast}(\rho)/m$ as functions of
density extracted from the ISGQR and IVGDR in $^{208}$Pb, respectively. The inset of
Fig.~\ref{dmrho} (a) shows the density dependence of the corresponding Pearson correlation
coefficient $r$ for $1/E_x^2$ vs $m_{s}^{\ast}(\rho)/m$ as well as $1/m_1$ vs
$m_{v}^{\ast}(\rho)/m$. The corresponding constraints on the isospin splitting coefficients
$\Delta m^*_1(\rho)$ and $\Delta m^*_3(\rho)$ as functions of density are shown in
Fig.~\ref{dmrho} (b). Also included in Fig.~\ref{dmrho} (b) are the $\Delta m^*_1(\rho_0)$
constraints obtained in Refs.~\cite{XuC10,LiXH15,LiBA13} as discussed earlier. Indeed,
one can see that all the $r$ values in the inset of Fig.~\ref{dmrho} (a) are larger than
$0.95$ for $0< \rho < 0.32$ fm$^{-3}$ that we are considering here, indicating the strong
linear correlation. Particularly, the strongest correlation appears at
$\rho \approx 0.19$ fm$^{-3}$ (with $r=0.97035$) for $1/E_x^2$ vs $m_{s}^{\ast}(\rho)/m$ while at
$\rho \approx 0.13$ fm$^{-3}$ (with $r=0.99961$) for $1/m_1$ vs $m_{v}^{\ast}(\rho)/m$.
It is seen that the $m_{s}^{\ast}(\rho)/m$ is generally larger than $m_{v}^{\ast}(\rho)/m$
and both $m_{s}^{\ast}(\rho)/m$ and $m_{v}^{\ast}(\rho)/m$ decease with density
but the latter exhibits a stronger density dependence, which leads to the
isospin splitting coefficients $\Delta m^*_1(\rho)$ and $\Delta m^*_3(\rho)$ increases
with density as observed in Fig.~\ref{dmrho} (b). It is interesting to see that the
third-order isospin splitting coefficient $\Delta m^*_3(\rho)$ is very small (about $0.05$
even at $\rho=0.32$ fm$^{-3}$) and can be approximately negligible. The
stronger isospin splitting of nucleon effective mass at higher densities may have
implications on the isospin effects in heavy ion collisions and neutrino emission in
neutron stars as mentioned earlier, and these are deserved further explorations in future.

Very recently, Roca-Maza \textit{et al.}~\cite{Roc15} pointed out that the
measured value of $\alpha_{\mathrm{D}}$ in $^{208}$Pb reported in Ref.~\cite{Tam11}
is contaminated by the non-resonant quasideuteron effect at higher energies.
The quasideuteron effect should be mainly due to correlated
neutron-proton pairs in nucleus and it overwhelms the IVGDR for the photon absorption
at higher energies beyond about $25$ MeV~\cite{Lep81}. In principle, the contribution
from the quasideuteron effect should be subtracted to directly compare the experimental
strength against the theoretical RPA calculations for the IVGDR.
In Ref.~\cite{Roc15} (and references therein), this contribution has been determined
and the electric dipole polarizability in $^{208}$Pb has been corrected to be
$19.6\pm0.6~\mathrm{fm}^3$.
Invoking the corrected value of $\alpha_{\mathrm{D}}$,
we derive a value of the EWSR of the IVGDR as $m_{1}=883.05\pm27.03
~\mathrm{MeV}\cdot\mathrm{fm}^2$ which is plotted as the cyan band
in Fig.~\ref{m1mv}.
Repeating the above analyses, one can further obtain the isovector effective
mass at $\rho_0$ as $m_{v,0}^{\ast}/m=0.80\pm0.03$ and the corrected linear
isospin splitting coefficient $\Delta m_1^{\ast}(\rho)$ as $\Delta m^*_1(\rho_0)=0.27\pm0.15$.
The obtained new value of the enhancement factor is $\kappa=0.197\pm0.037$, which is still
in good agreement with the results in Refs.~\cite{Tri08,Lip89}.
Again, we extract the density dependence of the isovector effective mass
$m_{v}^{\ast}$ and the isospin splitting coefficients $\Delta m^*_1(\rho)$ and
$\Delta m^*_3(\rho)$ by using the corrected value of $\alpha_{\mathrm{D}}$
in $^{208}$Pb due to the quasideuteron effect, and the results
are shown as the cyan bands in Fig.~\ref{dmrho}.
Overall, one can see that the correction on the
experimental value of $\alpha_{\mathrm{D}}$ in $^{208}$Pb from subtracting the contribution of
the quasideuteron effect does not affect the qualitative conclusions and
only leads to small corrections on quantitative results.

Furthermore, we have made similar analyses for the IVGDR of the semidouble-closed-shell
nucleus $^{68}$Ni. Using the measured centroid energy $E_{-1}=17.1\pm0.2$ MeV~\cite{Ros13}
as well as the electric dipole polarizability $\alpha_{\mathrm{D}}=3.88\pm0.31~\mathrm{fm}^{3}$
obtained in Ref.~\cite{Roc15} from a Lorenzian(-plus-Gaussian) extrapolation of
the measured GDR strength~\cite{Ros13} to the high-energy (low-energy) region, we extract
an isovector effective mass at $\rho_0$ as $m_{v,0}^{\ast}/m=0.81\pm0.11$, which is
in good agreement with the above results obtained from analyzing the data of $^{208}$Pb
although the uncertainty is larger. In addition, we would like to point out that
the present RPA calculations are not expected to reproduce the experimental
spreading width of the GDR, and this problem can be solved effectively by taking
into account the coupling to the collective low-lying (mainly surface) vibrations
or phonons~\cite{Niu14,Niu15,Lit15,Lyu15}. As discussed in Ref.~\cite{Roc15}, however, such an effect
beyond the mean-field approximation is not expected to significantly affect the
integral properties of the calculated strength that we are focusing on here.

\section{Conclusions}

Based on mean field calculations with Skyrme interactions, we have
demonstrated that the isoscalar and isovcetor effective masses at saturation
density, i.e., $m_{s,0}^{\ast}$ and $m_{v,0}^{\ast}$, can be well constrained
by the ISGQR excitation energy $E_x$ and the EWSR $m_1$ of the IVGDR in $^{208}$Pb,
respectively. In particular, invoking the experimental data for
$E_x$ in $^{208}$Pb, we have obtained the constraint
$m_{s,0}^{\ast}=0.91\pm0.05m$.
Meanwhile, combining the experimental IVGDR centroid energy and the electric
dipole polarizability $\alpha_{\mathrm{D}}=20.1\pm0.6$ in $^{208}$Pb, we have deduced a
value of $m_{1}=905.60\pm27.03~\mathrm{MeV}\cdot\mathrm{fm}^2$, and further
extracted a value of $m_{v,0}^{\ast}=0.77\pm0.03m$.
From the extracted $m_{s,0}^{\ast}$ and $m_{v,0}^{\ast}$, we have obtained a
constraint on the first-order (linear) isospin splitting coefficient of
nucleon effective mass, i.e., $\Delta m^*_1(\rho_0)=0.33\pm0.16$
which is in good agreement with the constraints extracted from global nucleon optical
potentials constrained by world data on nucleon-nucleus scattering~\cite{XuC10,LiXH15}
and is also consistent with the value obtained by analyzing the constraints
on the symmetry energy~\cite{LiBA13}.

Furthermore, we have constrained the isoscalar and isovcetor effective
masses as well as the isospin splitting of nucleon effective mass at other
densities by the similar analyses of the giant resonances in $^{208}$Pb.
Our results indicate that the isospin splitting of nucleon effective mass
increases with the density, and the third-order or higher-order isospin
splitting coefficients are negligibly small.

In addition, we have also investigated how our results change if the
recently corrected experimental value of $\alpha_{\mathrm{D}}=19.6\pm0.6$~fm$^3$
in $^{208}$Pb due to the quasideuteron effect is used. Our results
indicate that the corrected value leads to $m_{1}=883.1\pm27.0~\mathrm{MeV}\cdot\mathrm{fm}^2$,
$m_{v,0}^{\ast}=0.80\pm0.03~m$, and $\Delta m^*_1(\rho_0)=0.27\pm0.15$. Therefore,
the quasideuteron effect in $^{208}$Pb only plays a minor role on the
extractions of the isovector effective mass and the isospin splitting
coefficient of nucleon effective mass.
Our present work reveals for the first
time that the data on the giant resonances in $^{208}$Pb definitely favor
$m^{\ast}_{n}>m^{\ast}_{p}$ in neutron-rich matter, which sheds a
light upon understanding the isospin splitting of nucleon effective
mass in asymmetric nuclear matter.

\begin{acknowledgments}
We are grateful to Li-Gang Cao for helpful discussions on the
Skyrme-RPA code. This work was supported in part by the Major State Basic
Research Development Program (973 Program) in China under Contract Nos.
2013CB834405 and 2015CB856904, the NNSF of China under Grant Nos. 11275125
and 11135011, the ``Shu Guang" project supported by Shanghai Municipal
Education Commission and Shanghai Education Development
Foundation, the Program for Professor of Special Appointment (Eastern Scholar)
at Shanghai Institutions of Higher Learning, and the Science and Technology
Commission of Shanghai Municipality (11DZ2260700).
\end{acknowledgments}

\end{document}